\documentclass[a4paper,11pt]{article}
\pdfoutput=1
\usepackage{jheppub}
\usepackage{color,bm,comment,braket}
\usepackage[T1]{fontenc}
\usepackage{comment}
\newcommand{\del}{\partial}
\newcommand{\beq}{\begin{eqnarray}}
\newcommand{\eeq}{\end{eqnarray}}

\newcommand{\tr}{\mathop{\mathrm{tr}}}
\newcommand{\SU}{\text{SU}}
\newcommand{\U}{\text{U}}
\newcommand{\rmi}{\text{i}}
\newcommand{\rme}{\text{e}}
\newcommand{\rmd}{\text{d}}

\begin{document}

\title{
Phase diagram of QCD matter with magnetic field:  
domain-wall Skyrmion chain in chiral soliton lattice  
}
\author[a,b]{Minoru Eto}
\emailAdd{meto@sci.kj.yamagata-u.ac.jp}
\affiliation[a]{%
Department of Physics, Yamagata University, 
Kojirakawa-machi 1-4-12, Yamagata,
Yamagata 990-8560, Japan}
\affiliation[b]{Research and Education Center for Natural Sciences, Keio University, 4-1-1 Hiyoshi, Yokohama, Kanagawa 223-8521, Japan}

\author[c,d,b]{Kentaro Nishimura}
\affiliation[c]{
International Institute for Sustainability with Knotted Chiral Meta Matter(SKCM$^2$), Hiroshima University, 1-3-2 Kagamiyama, Higashi-Hiroshima, Hiroshima 739-8511, Japan
}
\affiliation[d]{KEK Theory Center, Tsukuba 305-0801, Japan}
\emailAdd{nishiken@post.kek.jp}

\author[e,b,c]{and Muneto Nitta}
\emailAdd{nitta@phys-h.keio.ac.jp}
\affiliation[e]{Department of Physics, Keio University, 4-1-1 Hiyoshi, Kanagawa 223-8521, Japan}

\abstract{
QCD matter in strong magnetic field exhibits a rich phase structure. In the presence of an external magnetic field, the chiral Lagrangian for two flavors is accompanied by the Wess-Zumino-Witten (WZW) term containing an anomalous coupling of the neutral pion $\pi_0$ to the magnetic field via the chiral anomaly. Due to this term, the ground state is inhomogeneous in the form of either chiral soliton lattice (CSL), an array of solitons in the direction of magnetic field, or domain-wall Skyrmion (DWSk) phase in which Skyrmions supported by $\pi_3[{\rm SU}(2)] \simeq {\mathbb Z}$ appear inside the solitons as topological lumps supported by $\pi_2(S^2) \simeq {\mathbb Z}$ in the effective worldvolume theory of the soliton. In this paper, we determine the phase boundary between the CSL and DWSk phases beyond the single-soliton approximation, within the leading order of chiral perturbation theory. To this end, we explore a domain-wall Skyrmion chain in multiple soliton configurations. First, we construct the effective theory of the CSL by the moduli approximation, and obtain the ${\mathbb C}P^1$ model or O(3) model, gauged by a background electromagnetic gauge field, with two kinds of topological terms coming from the WZW term: one is the topological lump charge in 2+1 dimensional worldvolume and the other is a topological term counting the soliton number. Topological lumps in the 2+1 dimensional worldvolume theory are superconducting rings and their sizes are constrained by the flux quantization condition. The negative energy condition of the lumps yields the phase boundary between the CSL and DWSk phases. We find that a large region inside the CSL is occupied by the DWSk phase, and that the CSL remains metastable in the DWSk phase in the vicinity of the phase boundary. 
}

\preprint{YGHP-23-05, KEK-TH-2573}

\maketitle

\section{Introduction}

The determination of matter phases stands as a pivotal challenge in modern physics. Quantum Chromodynamics (QCD) serves as the foundational theory of strong interactions, encapsulating descriptions of quarks, gluons, and hadrons (both baryons and mesons) as bound states of the aforementioned entities. Remarkably, the lattice QCD offers a comprehensive description of these bound states. The QCD phase diagram, especially under extreme conditions like high baryon density, pronounced magnetic fields, and rapid rotation, garners significant attention \cite{Fukushima:2010bq}. Such conditions are not merely of theoretical interest but pertain to real-world scenarios like the interior of neutron stars and phenomena observed in heavy-ion collisions. While lattice QCD is adept at addressing scenarios with zero baryon density, its extension to finite baryon density is hampered by the infamous sign problem. 
Contrastingly, in situations where chiral symmetry undergoes spontaneous breaking, the emergence of massless Nambu-Goldstone (NG) bosons or pions, which are dominant at low energy, is observed. This low-energy dynamics is aptly described by either the chiral Lagrangian or the chiral perturbation theory (ChPT) centered on the pionic degree of freedom. Importantly, this description is predominantly dictated by symmetries and only modulated by certain constants, including the pion's decay constant $f_{\pi}$ and quark masses $m_{\pi}$  \cite{Scherer:2012xha,Bogner:2009bt}.

\begin{figure}[tp]
  \begin{center}
   \includegraphics[width=15.0cm]{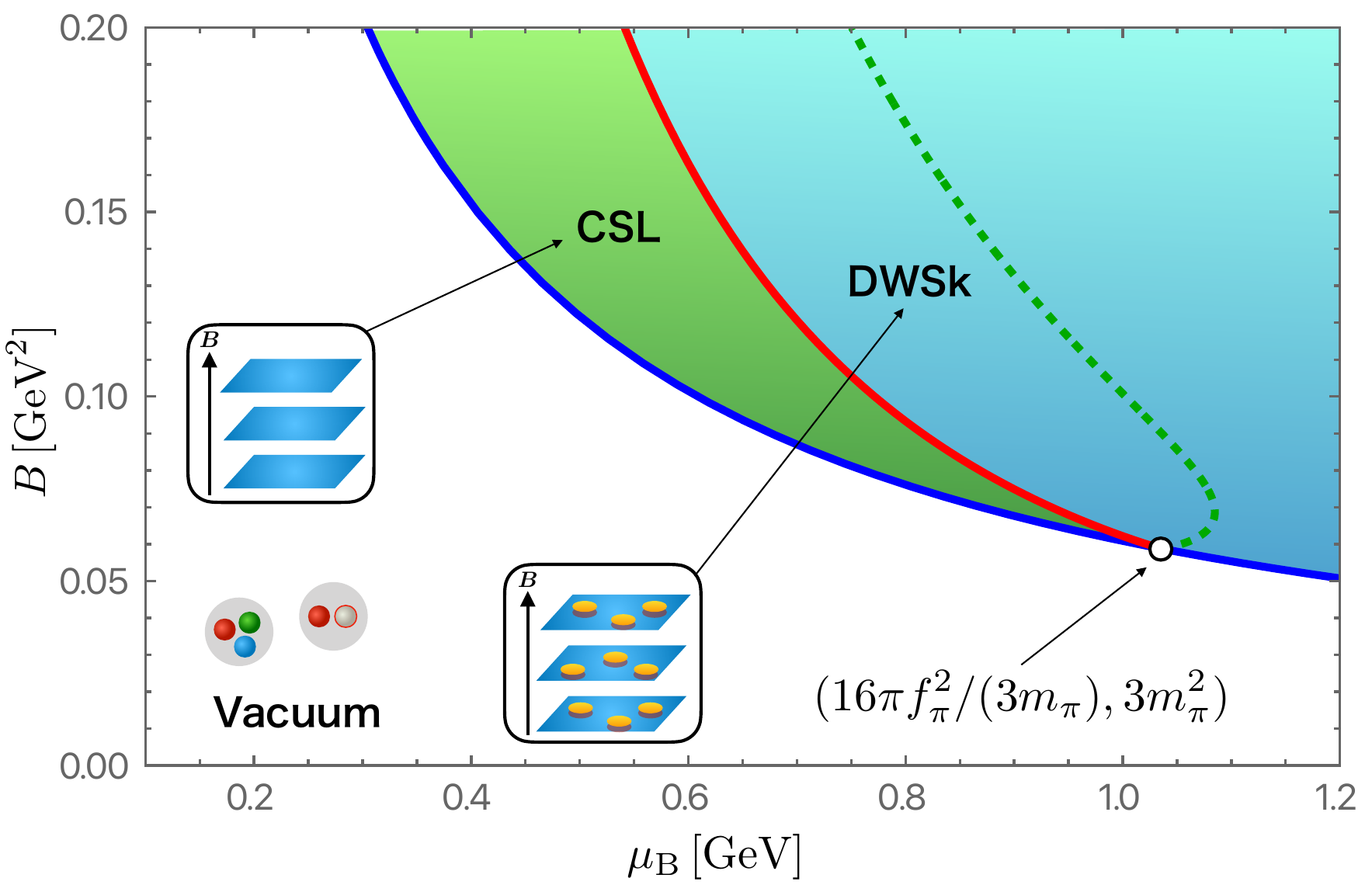}
  \end{center}
  \caption{
  Phase diagram of QCD matter with magnetic fields. 
  The blue curve denotes the boundary of QCD vacuum by Son and Stephanov in eq.~(\ref{eq:B_CSL}).
  The green dotted curve denotes the instability curve of the CSL by 
  the charged pion condensation 
  in eq.~(\ref{eq:instability-full}), given below, that asymptotically 
  reduces to eq.~(\ref{eq:instability}) 
  for large $B$. 
  The red curve is our new finding of the phase boundary between the CSL and DWSk phases  
  in eq.~(\ref{eq:phase-boundary}), given below, that behaves asymptotically as eq.~(\ref{eq:asympt}), given below.
  The blue, red and green dotted curves meet at the tricritical point in eq.~(\ref{eq:TCP}).
  The CSL configuration is a metastable state in the region between the red and green dotted curves.  
  }
  \label{fig:phase_diagram}
\end{figure}

As an extreme condition,
QCD in strong magnetic fields has received quite intense attention because of the interior of neutron stars 
and heavy-ion collisions.
In the presence of an external magnetic field,
the chiral Lagrangian is accompanied by 
the Wess-Zumino-Witten (WZW) term containing  
an anomalous coupling of the neutral pion $\pi_0$ to the magnetic field via the chiral anomaly 
\cite{Son:2004tq,Son:2007ny}
in terms of 
the Goldstone-Wilczek current \cite{Goldstone:1981kk,Witten:1983tw}.
It was determined to reproduce the so-called chiral separation effect  \cite{Vilenkin:1980fu,Son:2004tq,Metlitski:2005pr,Fukushima:2010bq,Landsteiner:2016led} in terms of the neutral pion $\pi_0$.
Then, at a finite baryon chemical potential $\mu_{\textrm{B}}$ under a sufficiently strong magnetic field $B$, if the inequality 
\begin{gather}
    B  
    \geq B_{\rm CSL} =
    \frac{16\pi m_{\pi}f_{\pi}^2}{e \mu_{\textrm{B}}}  
    \mbox{: the blue curve in fig.~\ref{fig:phase_diagram}}
    \label{eq:B_CSL} \, 
\end{gather}
holds, 
the ground state of QCD with two flavors (up and down quarks) becomes inhomogenous  
in the form of a chiral soliton lattice (CSL) 
consisting of a stack of 
domain walls or solitons 
carrying a baryon number 
\cite{Son:2007ny,Eto:2012qd,Brauner:2016pko}.\footnote{
CSLs manifest in diverse scenarios within QCD. Examples include situations under rapid rotation~\cite{Huang:2017pqe,Nishimura:2020odq,Chen:2021aiq,Eto:2021gyy,Eto:2023tuu} and due to thermal fluctuations~\cite{Brauner:2017uiu,Brauner:2017mui,Brauner:2021sci,Brauner:2023ort}. Investigations into the quantum nucleation of CSLs can be found in~\cite{Eto:2022lhu,Higaki:2022gnw}, and discussions on quasicrystals are provided in~\cite{Qiu:2023guy}. The potential interplay between Skyrmion crystals at zero magnetic field and the CSL phase has been explored in Refs.~\cite{Kawaguchi:2018fpi,Chen:2021vou,Chen:2023jbq}.
}

However, Brauner and Yamamoto found that 
such a CSL state is unstable against 
a charged pion condensation 
in a region of higher density and/or stronger magnetic field \cite{Brauner:2016pko}. 
The asymptotic expression of the instability curve 
at large $B$ is
\begin{eqnarray}
    B_{\rm CPC} 
    \sim \frac{16 \pi^4 f_{\pi}^4}{\mu_{\rm B}^2} 
    \mbox{: the green dotted curve (at large $B$) in fig.~\ref{fig:phase_diagram}}
    \label{eq:instability}
\end{eqnarray}
above which the CSL is unstable, 
where ``CPC" denotes the charged pion condensation.
The full expression of the boundary is given 
in eq.~(\ref{eq:instability-full}),
below, 
and is denoted by the green dotted curve 
in fig.~\ref{fig:phase_diagram}.
In Ref.~\cite{Evans:2022hwr},
 an Abrikosov's vortex lattice was proposed as a consequence of the charged pion condensation 
(see also a recent paper \cite{Evans:2023hms}). 
This instability curve ends 
at the tricritical point:
\begin{eqnarray}
    (\mu_{\rm c},B_{\rm c}) 
    &=& 
    \left(\frac{16\pi f_{\pi}^2}{3m_{\pi}}, 
    \frac{3m_{\pi}^2}{e} \right) \nonumber\\
    &\approx& 
    \left(1.03 \;\; {\rm GeV}, 
    0.06 {\rm GeV}^2\sim 1.0\times 10^{19} {\rm G}\right)  \mbox{: the white dot in fig.~\ref{fig:phase_diagram}},
    \label{eq:TCP}
\end{eqnarray}
with the vacuum values of the physical quantities $f_{\pi}\approx 93\, \textrm{MeV}$ and $m_{\pi}\approx 140\, \textrm{MeV}$.

In our previous paper \cite{Eto:2023lyo}, 
we proposed that 
there is the domain-wall Skyrmion (DWSk) phase in the region inside the CSL,
\begin{eqnarray}
\mu_{\rm B} \geq \mu_{\rm c}
  \label{eq:negative}
\end{eqnarray}
  with 
$\mu_{\rm c}$ in eq.~(\ref{eq:TCP}), 
in which Skyrmions are created on top of the solitons 
in the ground state.
To show this,
the effective world-volume theory on a single soliton was constructed as an O(3) sigma model 
or the ${\mathbb C}P^1$ model with topological terms induced from the WZW term. Then, there appear 
topological lumps (or baby Skyrmions) 
supported by $\pi_2({\mathbb C}P^1)\simeq {\mathbb Z}$
on the world volume, 
corresponding to 3+1 dimensional Skyrmions supported by 
$\pi_3[{\rm SU}(2)]\simeq {\mathbb Z}$
in the bulk point of view. 
Such a composite state of a domain wall and Skyrmions are called 
domain-wall Skyrmions.\footnote{
Domain-wall Skyrmions were initially introduced in the context of field theory, both in 3+1 dimensions~\cite{Nitta:2012wi,Nitta:2012rq,Gudnason:2014nba,Gudnason:2014hsa,Eto:2015uqa,Nitta:2022ahj} and in 2+1 dimensions~\cite{Nitta:2012xq,Kobayashi:2013ju,Jennings:2013aea}. In the realm of condensed matter physics, the 2+1 dimensional variant has been both theoretically investigated~\cite{PhysRevB.99.184412,KBRBSK,Ross:2022vsa,Amari:2023gqv,Amari:2023bmx,PhysRevB.102.094402,Kim:2017lsi,Lee:2022rxi} and experimentally observed in chiral magnets~\cite{Nagase:2020imn,Yang:2021}. 
The term ``domain-wall Skyrmions'' can be traced back to its initial use in ref.~\cite{Eto:2005cc}, where it described Yang-Mills instantons situated within a domain wall in 4+1 dimensions. In this setting, these entities are represented as Skyrmions in the effective 3+1 dimensional domain-wall theory. However, a more appropriate designation for them might be ``domain-wall instantons.''
}
However, we used a single-soliton approximation, 
considering 
the domain-wall Skyrmion on a single soliton 
\cite{Eto:2023lyo}. In other words,
we assumed that the solitons are well separated, 
and this assumption can be justified only at the phase boundary 
between the QCD vacuum and the CSL phase 
in eq.~(\ref{eq:B_CSL}), 
namely at the tricritical point in eq.~(\ref{eq:TCP}). 
On the other hand, the instability curve of the CSL 
due to a charged pion condensation 
also ends at the same point in eq.~(\ref{eq:TCP}) 
\cite{Brauner:2016pko}. 
Therefore, a natural question that arises is the compatibility between
the DWSk phase and instability curve.

In this paper, we determine  
the phase boundary between 
the CSL and DWSk phases 
beyond 
the single-soliton approximation,  
in which the boundary was the straight vertical line in eq.~(\ref{eq:negative}) 
ending on the tricritical point 
represented by the white dot in fig.~\ref{fig:phase_diagram}.
To this end,
we explore domain-wall Skyrmion chains 
in multiple soliton configurations.
A similar domain-wall skyrmion chain has been also studied in chiral magnets \cite{Amari:2023bmx}.
As is well known, the CSL configuration is  analytically 
given by the elliptic function. 
We construct the effective theory 
of the CSL 
by the moduli approximation \cite{Manton:1981mp,Eto:2006pg,Eto:2006uw}, 
in which we promote the $\mathbb{C}P^1$ moduli of the CSL
to  fields depending 
on the worldvolume $(x^0,x^1,x^2)$ and integrate 
over one period of the lattice in the codimensional direction $x^3$. 
We obtain the ${\mathbb C}P^1$ model or O(3) model with two kinds of topological terms coming from 
the WZW term: one is topological lump charge 
responsible for 
$\pi_2(S^2) \simeq {\mathbb Z}$ in 2+1 dimensional worldvolume 
and the other is a topological term 
counting the soliton number.
We then construct lumps in the 
 2+1 dimensional worldvolume theory.
 Since the electromagnetic U(1) gauge symmetry 
 is spontaneously broken around a ring surrounding the lump, 
 the lump can be regarded as a superconducting ring. 
 Then its size modulus is fixed
 by the flux quantization condition of the superconducting ring, 
 enhancing its stability. 
 The lumps in the soliton worldvolume correspond to Skyrmions in the bulk, 
 Skyrmions periodically sit on each soliton in the CSL, 
and thus the configuration is a domain-wall Skyrmion chain. 
 The condition that a lump has negative energy 
 yields the phase boundary between 
 the CSL and DWSk phases 
 denoted by the red curve in fig.~\ref{fig:phase_diagram}.
 In the strong magnetic field limit,  
 the phase boundary asymptotically behaves as 
\begin{eqnarray}
    B_{\rm c} \sim \frac{4 \pi^3 f_{\pi}^4}{\mu_{\rm B}^2} 
    = \frac{1}{4\pi} B_{\rm CPC}.
\end{eqnarray}
 The important is that the boundary curve 
 has the lower critial chemical potential $\mu_{\rm B}$ and 
 lower critical magnetic field $B$ than those of the instability curve 
 (the green dotted curve in fig.~\ref{fig:phase_diagram})
 of the CSL in eq.~(\ref{eq:instability}). 
 Therefore, the CSL state remains metastable in the region between the red and green dotted curves.

This paper is organized as follows.
In sec.~\ref{sec:CSL} we present a CSL in the strong magnetic field.
In sec.~\ref{sec:eff-th} we construct 
the effective worldvolume theory of a one period of the CSL, by the moduli approximation.
In sec.~\ref{sec:DWSk-chain} we construct topological lumps in the soliton's worldvolume theory and determine the phase boundary between the CSL and DWSk phases.
Sec.~\ref{sec:summary} is devoted to a summary and discussion.

\section{Chiral soliton lattice in strong magnetic field} \label{sec:CSL}
We focus on the phase where chiral symmetry is spontaneously broken.
The effective field theory of pions, known as ChPT, can describe the low-energy dynamics.
The pion fields are represented by a $2\times 2$ unitary matrix,
\begin{equation}
    \Sigma = \rme^{\rmi \tau_a \chi_a} \,,
\end{equation}
where $\tau_a$ (with \(a = 1, 2, 3\)) are the Pauli matrices,
normalized as \(\tr(\tau_a \tau_b) = 2 \delta_{ab}\).
The field \(\Sigma\) transforms under \(\SU(2)_{\mathrm{L}} \times \SU(2)_{\mathrm{R}}\) chiral symmetry as
\begin{equation}
    \Sigma \rightarrow L \Sigma R^{\dagger} \,,
\end{equation}
where both \(L\) and \(R\) are \(2 \times 2\) unitary matrices.
Then, the effective Lagrangian at the ${\cal O}(p^2)$ order is ($\mu=0,\cdots,3$)
\begin{gather}
    \mathcal{L}_{\textrm{ChPT}}
    = \frac{f_{\pi}^2}{4} \tr \left(D_{\mu}\Sigma D^{\mu}\Sigma^{\dag} \right) 
    - \frac{f_{\pi}^2m_{\pi}^2}{4}
    \tr 
    (2 {\bf 1}-\Sigma-\Sigma^{\dag}) \label{ChPT_with_B} \,,
\end{gather}
where $f_{\pi}$ and $m_{\pi}$ are pion's decay constant and mass, respectively, and 
$D_{\mu}$ is a covariant derivative defined by
\begin{gather}
    D_{\mu}\Sigma
    \equiv \del_{\mu}\Sigma
    + \rmi e A_{\mu} [Q, \Sigma] \label{def_cov_del} \,, \\
    Q = \frac{1}{6}\bm{1} + \frac{1}{2}\tau_3.
    \,,
\end{gather}
where $Q$ is a matrix of the electric charge of quarks.
The \( \U(1)_{\mathrm{EM}} \) transformation is given by
\begin{equation}
    \Sigma \to \rme^{\mathrm{i}\lambda\frac{\tau_3}{2}} \Sigma \rme^{-\mathrm{i}\lambda\frac{\tau_3}{2}} \quad \text{and} \quad A_\mu \to A_\mu - \frac{1}{e} \partial_\mu \lambda.
\end{equation}
The external \( \U(1)_{\mathrm{B}} \) gauge field \( A^{\mathrm{B}}_{\mu} \) can couple to \(\Sigma\) via the Goldstone-Wilczek current \cite{Goldstone:1981kk,Witten:1983tw}.
The conserved and gauge-invariant baryon current in the external magnetic field is detailed in refs.~\cite{Goldstone:1981kk,Son:2007ny}:
\begin{equation}
    j_{\mathrm{GW}}^{\mu} = -\frac{\epsilon^{\mu \nu \alpha \beta}}{24\pi^2} \mathrm{tr} \left( L_{\nu}L_{\alpha}L_{\beta} - 3\mathrm{i} e \partial_{\nu} \left[ A_{\alpha} Q(L_{\beta} + R_{\beta}) \right] \right) \,,
    \label{eq:GW}
\end{equation}
with \( A^{\mathrm{B}}_{\mu} = (\mu_{\mathrm{B}}, \bm{0}) \), and introducing the notations \( L_{\mu} \equiv \Sigma \partial_{\mu} \Sigma^\dagger \) and \( R_{\mu} \equiv \partial_{\mu} \Sigma^\dagger \Sigma \).

The effective Lagrangian that couples to \( A^{\mathrm{B}}_{\mu} \) is expressed as
\begin{equation}
    \mathcal{L}_{\mathrm{WZW}} = -A^{\mathrm{B}}_{\mu} j_{\mathrm{GW}}^{\mu},
    \label{effective_Lagrangian_GW}
\end{equation}
which is recognized as the WZW term \cite{Son:2004tq,Son:2007ny}. Thus, the total Lagrangian is
\begin{equation}
    \mathcal{L} = \mathcal{L}_{\mathrm{ChPT}} + \mathcal{L}_{\mathrm{WZW}}.
\end{equation}

An important observation is warranted at this point.
In order to formulate an effective Lagrangian,
we adopt a modification of the standard power counting scheme of ChPT as presented in ref.~\cite{Brauner:2021sci}:
\begin{equation}
    \partial_{\mu} \,,
    m_{\pi} \,,
    A_{\mu} = \mathcal{O}(p^1) \,, \quad
    A^{\mathrm{B}}_{\mu} = \mathcal{O}(p^{-1}).
\end{equation}
In this power-counting scheme,
eq.~(\ref{effective_Lagrangian_GW}) is of order $\mathcal{O}(p^2)$ and is consistent with eq.~(\ref{ChPT_with_B}).
It is significant to note that $\mu_{\mathrm{B}}$ only manifests in the WZW term of eq.~(\ref{effective_Lagrangian_GW}),
which allows us to attribute a negative power counting to $\mu_{\mathrm{B}}$.
The effective field theory up to $\mathcal{O}(p^2)$ must incorporate both terms in eq.~(\ref{eq:GW}).
However, previous studies on the CSLs have not taken into account the first term in eq.~(\ref{eq:GW}). 

We emphasize that the inclusion of an $\mathcal{O}(p^4)$ term,
such as the Skyrme term and the chiral anomaly term (which includes $\pi_0\bm{E}\cdot \bm{B}$),
is not essential for our results.
As a result, our analysis maintains its model-independece.
Moreover, it should be noted that at the leading order,
the gauge field is nondynamical due to its kinetic term being of order $\mathcal{O}(p^4)$.

We note that our effective theory admits a parallel stack of the sine-Gordon soliton expanding perpendicular to the external magnetic field,
which is called the chiral soliton lattice.
This state is stable under a sufficiently large magnetic field, as shown in \cite{Son:2007ny}.
If we consider the case of 
no charged pions 
$\Sigma_0 = \rme^{\rmi \tau_3\chi_3}$, the effective Lagrangian  reduces to 
\begin{eqnarray}
\mathcal{L}
= \frac{f_{\pi}^2}{2}(\del_{\mu}\chi_3)^2
    - f_{\pi}^2m_{\pi}^2(1-\textrm{cos}\chi_3)
    + \frac{e\mu_{\textrm{B}}}{4\pi^2}\bm{B}\cdot \bm{\nabla}\chi_3 \,. \label{eq:lagrangian_without_charged_pions}
\end{eqnarray}

The ordinary QCD vacuum corresponds to $\chi_3=0$.
However, the third term in eq.~(\ref{eq:lagrangian_without_charged_pions}) modifies the ground state of QCD at finite $\mu_{\textrm{B}}$ and $\bm{B}$.
The anticipated time-independent neutral pion background $\chi_3$ is obtained by minimization of the energy functional.
The static Hamiltonian depending only on the $z$ coordinate is given by
\begin{gather}
    \mathcal{H}
    = 
    \frac{f_{\pi}^2}{2}(\del_z\chi_3)^2
    + f_{\pi}^2m_{\pi}^2(1-\cos \chi_3)
    -\frac{e\mu_{\textrm{B}}B}{4\pi^2}\del_z\chi_3 \,. \label{eq:sine-Gordon_Hamiltonian_with_topo}
\end{gather}
Without loss of generality, we will orient the uniform external magnetic field along the $z$-axis; $\bm{B}=(0,0,B)$.
We note that eq.~(\ref{eq:sine-Gordon_Hamiltonian_with_topo}) have the first derivative term proportional to $\del_z\chi_3$.
Then, the configuration of the ground state will have a nontrivial $z$-dependence.
In order to determine the static configuration of $\chi_3$,
let us solve the EOM of eq.~(\ref{eq:lagrangian_without_charged_pions}).
The equation of motion for such a one-dimensional configuration $\chi_3(z)$ then reads
\begin{gather}
    \del_z^2\chi_3 = m_{\pi}^2\sin \chi_3 \label{eq:Jacobi-amplitude} \,.
\end{gather}
which can be analytically solved by the elliptic functions:
\begin{gather}
    \chi_3^{\textrm{CSL}} = 2\textrm{am}\left(\frac{m_{\pi}z}{\kappa},\kappa \right) + \pi \,.\label{eq:Jacobi-amplitude-sol} 
\end{gather}
with a real constant $\kappa$ ($0\leq\kappa \leq1$) called the elliptic modulus. 
This solution is a lattice state of the $\pi_0$ $(=f_\pi \chi_3)$ meson with a period $\ell$.
The period $\ell$ satisfies the following equations:
\begin{gather}
    \chi_3^{\textrm{CSL}}(z+\ell) = \chi_3^{\textrm{CSL}}(z) + 2\pi \,, \\
    \ell = \frac{2\kappa K(\kappa)}{m_{\pi}} \,,
\end{gather}
with the complete elliptic integral  of the first kind $K(\kappa)$.
Substituting eq.~(\ref{eq:Jacobi-amplitude-sol}) into eq.~(\ref{eq:sine-Gordon_Hamiltonian_with_topo}) and integrating it between one period $\ell$,
we get 
the tension of a single soliton inside the CSL, that is the energy density per unit area integrated over one period 
\begin{gather}
    \mathcal{E} = \int_{0}^{\ell} \rmd z\, \mathcal{H}
    = 4m_{\pi}f_{\pi}^2\left[
    \frac{2E(\kappa)}{\kappa} + \left(\kappa-\frac{1}{\kappa} \right)K(\kappa)
    \right] \,,
    \label{eq:energy_denstiy_csl}
\end{gather}
with the complete elliptic integral  of the second kind $E(\kappa)$.
Minimizing the energy density per unit length $\mathcal{E}/\ell$ with respect to $k$ gives me the following condition:
\begin{align}
    \frac{E(\kappa)}{\kappa} &= \frac{e\mu_{\text{B}}B}{16\pi m_{\pi}f_{\pi}^2} \label{eq:minimization_condition} .
\end{align}
Since the left-hand side of Eq.~\eqref{eq:minimization_condition} is bounded from below as $E(\kappa)/\kappa\ge 1\, (0\le \kappa \le 1)$,
the CSL solution exists if and only if the following condition is satisfied \cite{Son:2007ny,Brauner:2016pko}:
\begin{align}
    B_{\text{CSL}} &= \frac{16\pi m_{\pi}f_{\pi}^2}{e\mu_{\text{B}}} ,
\end{align}
denoted by the blue curve in fig.~\ref{fig:phase_diagram}.
Inserting the minimization condition \eqref{eq:minimization_condition} into Eq.~\eqref{eq:energy_denstiy_csl},
we evaluate the energy density at the optimized $k$ as:
\begin{align}
    \mathcal{E} &= 4m_{\pi}f_{\pi}^2\left(\kappa-\frac{1}{\kappa} \right)K(\kappa) < 0,
\end{align}
which is lower than that of the QCD vacuum.
Therefore, the CSL is energetically more stable than the QCD vacuum.

\section{Effective worldvoume theory of soliton in chiral soliton lattice}
\label{sec:eff-th}

The preceding section concentrated exclusively on the $\pi_0$ meson.
General solutions that encompass charged pions can be derived from $\Sigma_0$ through an $\SU(2)_{\text{V}}$ transformation,
\begin{align}
    \Sigma = g\Sigma_0 g^{\dag} = \exp(\rmi \chi_3^{\text{CSL}} g \tau_3 g^{\dag}) \label{general_sol}\,,
\end{align}
where $g$ represents an $\SU(2)$ matrix.
It is clear that $\Sigma_0$ is invariant under $\SU(2)_{\text{V}}$ when $g=\rme^{\rmi \tau_3\theta}$.
Consequently, each soliton possesses moduli originated from the spontaneous symmetry breaking $\SU(2)_{\text{V}}\to \U(1)_3$ in the vicinity of the soliton:
\begin{align}
    \mathcal{M} \cong \frac{\SU(2)_{\text{V}}}{\U(1)_3} \cong {\mathbb C}P^1 \cong S^2\,.
\end{align}
Such a sine-Gordon soliton carrying non-Abelian ${\mathbb C}P^1$ moduli is 
called a non-Abelian sine-Gordon soliton \cite{Nitta:2014rxa,Eto:2015uqa} 
(see also refs.~\cite{Nitta:2015mma,Nitta:2015mxa,Nitta:2022ahj}). 
However, unlike these references, our solitons are nontopological because they are not protected by topology and are rather stabilized by the WZW term.

For subsequent discussions,
we characterize the $\mathbb{C}P^1$ moduli using the homogeneous coordinates $\phi \in \mathbb{C}^2$ of $\mathbb{C}P^1$,
which fulfill the relations \cite{Eto:2015uqa}
\begin{align}
    \phi^{\dag}\phi = 1, \quad
    g \tau_3 g^{\dag} = 2 \phi \phi^{\dag} - \bm{1}_2\,.
\end{align}
In the context of $\phi$,
eq.~\eqref{general_sol} can be recast as
\begin{align}
    \Sigma &= \exp(2\rmi 
    \chi_3^{\rm CSL}
    \phi \phi^{\dag})u^{-1} = [{\bm 1}_2 + (u^2-1)\phi \phi^{\dag}]u^{-1} \label{general_sol_Sigma}\,, \\
    u &\equiv \rme^{\rmi \chi_3^{\rm CSL}} =\exp\left(2\,\rmi\,\textrm{am}\left(\frac{m_{\pi}z}{\kappa},\kappa \right) + \pi \rmi\right)\,.
\end{align}
Given that the moduli space is $S^2$,
we can also employ the real three-component vector ${\bm n}$ defined as
\begin{gather}
    n_a = \phi^{\dag}\tau_a\phi \,,
\end{gather}
to describe this space.
The $\pi^0$ CSL in eq.~(\ref{eq:Jacobi-amplitude-sol}) corresponds to $n_3 = 1$.

Let us construct the low-energy effective field theory of the CSL based on the moduli approximation \cite{Manton:1981mp,Eto:2006pg,Eto:2006uw}.
In the following, we will promote the moduli parameter $\phi$ as the fields on the $2+1$-dimensional soliton's world volume.
We first calculate the effective action coming from eq.~(\ref{ChPT_with_B}).
Substituting eq.~(\ref{general_sol_Sigma}) into eq.~(\ref{ChPT_with_B}),
we get
\begin{align}
    \mathcal{L}_{\textrm{ChPT}}
    &= \frac{f_{\pi}^2}{2} |1-u^2|^2 [(\phi^{\dag}\del_{\alpha}\phi)^2+\del_{\alpha}\phi^{\dag}\del^{\alpha}\phi] \notag \\ 
    &+ \frac{e}{2}f_{\pi}^2A^{\alpha}|u^2-1|^2 
    \left[
    \phi^{\dag}\tau_3\phi \cdot \phi^{\dag}\del_{\alpha}\phi
    + \frac{1}{2}(\del_{\alpha}\phi^{\dag}\tau_3\phi - \phi^{\dag}\tau_3\del_{\alpha}\phi)
    \right] \notag \\
    &- \frac{e^2 f_{\pi}^2}{8}A^2 |u^2-1|^2 \left[-1 + (\phi^{\dag}\tau_3\phi)^2 \right] \notag \\
    &- \frac{f_{\pi}^2}{2} (\del_z
    \chi_3^{\rm CSL}
    )^2 - f_{\pi}^2m_{\pi}^2(1-\textrm{cos}
   \chi_3^{\rm CSL}) \,,
    \label{eq:ChPT_with_B2}
\end{align}
where $x^\alpha$ ($\alpha =0,1,2$) are world-volume coordinates.
Integrating over $z$, the effective action stemming from eq.~(\ref{eq:ChPT_with_B2}) can be calculated as
\begin{align}
    \int \rmd z\, \mathcal{L}_{\textrm{ChPT}}
    & = {\cal C}(\kappa)
    [(\phi^{\dag}D_{\alpha}\phi)^2
    +D_{\alpha}\phi^{\dag}D^{\alpha}\phi]
    - \mathcal{E}
    \label{int_z_ChPT} \,,
\end{align}
with the K\"ahler class ${\cal C}(\kappa)$ defined by
\begin{align}
  {\cal C}(\kappa) \equiv \frac{16f_{\pi}^2}{3m_{\pi}}
    \frac{(2-\kappa^2)E(\kappa)-2(1-\kappa^2)K(\kappa)}{\kappa^3}
    \label{eq:fk}
\end{align}
and the tension 
$\mathcal{E}$ of a signle soliton in eq.~(\ref{eq:energy_denstiy_csl}), 
where we have used the integrals,
\begin{gather}
    \int_{0}^{\frac{2\kappa K(\kappa)}{m_{\pi}}} \rmd z\, |1-u^2|^2
    = \frac{16}{3m_{\pi}}
    \frac{
    2(2-\kappa^2)E(\kappa) - 4(1-\kappa^2)K(\kappa)
    }{\kappa^3} \,,
\end{gather}
and the contribution of the gauge field can be summarized into the covariant derivative:
\begin{gather}
    D_{\alpha}\phi
    = \left(
    \del_{\alpha}+\rmi e A_{\alpha}\frac{\tau_3}{2}
    \right)\phi \,.
\end{gather}
The first term in eq.~(\ref{int_z_ChPT}) is the kinetic term of $\phi$, being equivalent to the gauged $\mathbb{C}P^1$ model. 
The second term in eq.~(\ref{int_z_ChPT})  is the minus 
tension of each soliton, that is
the energy density of the CSL in one period $\ell$.

We next calculate the effective action coming from the WZW term in eq.~(\ref{effective_Lagrangian_GW}).
The contribution of
the first term in eq.~(\ref{eq:GW}) 
to 
the WZW term in eq.~(\ref{effective_Lagrangian_GW}) 
can be expressed as $\mu_{\textrm{B}}{\cal B}$.
Here, we refer to the Skyrmoin charge density as
\begin{gather}
    {\cal B} = \frac{-1}{24\pi^2}\epsilon^{ijk} \tr(L_{i}L_{j}L_{k}) \,,
\end{gather}
which can be factorized as
\begin{eqnarray}
    {\cal B} = -\frac{1}{2\pi}(u-u^{-1})^2\del_z\chi^{\textrm{CSL}}_3 q(x,y) \,,
\end{eqnarray}
where the $\mathbb{C}P^1$ lump topological charge density is defined as
\begin{eqnarray}
    q(x,y) \equiv
     -\frac{\rmi}{2\pi}\epsilon^{ij}\del_i\phi^{\dag}\del_j\phi
     = \frac{1}{8\pi} \epsilon^{ij}{\bm n}\cdot (\partial_i {\bm n}\times\partial_j{\bm n}) \,.
\end{eqnarray}
Integration of $q$ over $x$ and $y$ gives the quantized lump charge $k$, being associated with $\pi_2({\mathbb C}P^1)$ :
\begin{gather}
  k \equiv \int \rmd^2 x\, q 
  \in \pi_2({\mathbb C}P^1) \simeq \mathbb{Z} \,.
\end{gather}
Integrating ${\cal B}$ from $0$ to $\ell$, we get
\begin{eqnarray}
    \int^{\ell}_{0}dz\, {\cal B} = 2 q(x,y) \,,
    \label{eq:baryon-density}
\end{eqnarray}
where we have used 
\begin{align}
    -\int_{0}^{\ell}\rmd z\, \frac{1}{2\pi}(u-u^{-1})^2\del_z\chi^{\textrm{CSL}}_3
    = \frac{16}{\pi}\int_{0}^{2K(\kappa)}\rmd \bar{z}\,
    \textrm{sn}(\bar{z},\kappa)^2\textrm{cn}(\bar{z},\kappa)^2\textrm{dn}(\bar{z},\kappa)
    = 2 \,. \label{eq:integral_emGW_current}
\end{align}
Integrating eq.~(\ref{eq:baryon-density}) over the $xy$ plane, we find the barynon number $b$ per one period: 
\begin{equation}
   b = 2k \in \pi_3[{\rm SU}(2)].\label{eq:pi2-pi3}
\end{equation}
We thus have seen that one lump on one soliton corresponds to two Skyrmions (baryons) in the bulk.
This one-to-two correspondence is in contrast to 
the domain-wall Skyrmions in QCD under rapid rotation 
\cite{Eto:2023tuu}, 
in which case one lump on a soliton corresponds to one Skyrmion.

The contribution of the second term in eq.~(\ref{eq:GW}) 
to the WZW term in eq.~(\ref{effective_Lagrangian_GW})
can be divided into two terms as follows:
\begin{align}
    \frac{\rmi e\mu_{\textrm{B}}}{16\pi^2} \epsilon^{0ijk} \del_i[A_j\tr(\tau_3L_k+\tau_3R_k)]
    &= \frac{\rmi e\mu_{\textrm{B}}}{16\pi^2}\epsilon^{0ijk} \del_iA_j \tr(\tau_3L_k+\tau_3R_k) \notag \\
    &+ \frac{\rmi e\mu_{\textrm{B}}}{16\pi^2}\epsilon^{0ijk} A_j \tr \tau_3 (\del_i\Sigma \del_k\Sigma^{\dag} + \del_k\Sigma^{\dag}\del_i\Sigma) \label{WZW_gauged_term} \,.
\end{align}
We consider the uniform external magnetic field along the $z$-axis,
$\bm{B}=(0,0,B)$.
Then, the first term in eq.~(\ref{WZW_gauged_term}) becomes
\begin{gather}
    -\frac{\rmi e\mu_{\textrm{B}}}{16\pi^2}B \tr \tau_3(L_3+R_3) \label{domain-wall_charge} \,.
\end{gather}
In terms of the projection operator 
$P\equiv \phi \phi^{\dag}$ 
satisfying $P^2 = P$,
$R_k$ and $L_k$ can be expressed as
\begin{eqnarray}
    &&L_k = (1-2P) \rmi \del_k \chi_3^{\rm CSL} + (u^{-2}-1)\del_kP + |u^2-1|^2P\del_kP \,, \\
    &&R_k = (1-2P) \rmi \del_k \chi_3^{\rm CSL} + (u^{-2}-1)\del_kP + |u^2-1|^2\del_kP \cdot P \,.
\end{eqnarray}
Since $\phi$ does not depend on $z$,
the second and third terms in $L_3$ and $R_3$ vanish.
Therefore, eq.~(\ref{domain-wall_charge}) becomes
\begin{gather}
    -\frac{e \mu_{\textrm{B}}B}{4\pi^2} (\phi^{\dag}\tau_3\phi) \del_3\chi^{\textrm{CSL}}_3 \,,
\end{gather}
and integrating over $z$, we get
\begin{gather}
    \int_{0}^{\ell} \rmd z\,
    \frac{\rmi e\mu_{\textrm{B}}}{16\pi^2}\epsilon^{0ijk} \del_iA_j \tr(\tau_3L_k+\tau_3R_k)
    = -\frac{e \mu_{\textrm{B}}B}{2\pi} \phi^{\dag}\tau_3\phi \label{WZW_first_term} \,,
\end{gather}
where we have used the boundary condition of 
$\chi_3^{\rm CSL}$ for a single soliton, $\chi_3^{\rm CSL}(\ell)-\chi_3^{\rm CSL}(0) = 2\pi$.
We next calculate the second term in eq.~(\ref{WZW_gauged_term}).
Substituting eq.~(\ref{general_sol_Sigma}) into $\del_i\Sigma \del_k\Sigma^{\dag}$ and $\del_k\Sigma^{\dag} \del_i\Sigma$, these two quantities can be calculated as
\begin{align}
    \del_i\Sigma \del_k\Sigma^{\dag}
    &= [(1-u^{-2}) -(u^2 - u^{-2})P] \rmi \del_i
    \chi_3^{\rm CSL} \del_kP \nonumber\\
 &- [(1-u^{2})+(u^2 - u^{-2})P] \rmi \del_k
 \chi_3^{\rm CSL} \del_iP
    + |1-u^2|^2 \del_iP \del_kP \,, \\
    \del_k\Sigma^{\dag} \del_i\Sigma
    &= [(1-u^{-2}) -(u^2 - u^{-2})P] \rmi \del_i
    \chi_3^{\rm CSL} \del_kP \nonumber\\
 &- [(1-u^{2})+(u^2 - u^{-2})P] \rmi \del_k
 \chi_3^{\rm CSL} \del_iP
    + |1-u^2|^2 \del_kP \del_iP \,.
\end{align}
Hence, we can represent $\epsilon^{0ijk}\tr(\del_i\Sigma \del_k\Sigma^{\dag} + \del_k\Sigma^{\dag} \del_i\Sigma)$ in terms of $u$ and $\phi$ as follows:
\begin{align}
    &\epsilon^{0ijk}\tr \tau_3(\del_i\Sigma \del_k\Sigma^{\dag} + \del_k\Sigma^{\dag} \del_i\Sigma) \notag \\
    &= \epsilon^{0ijk}\tr \tau_3 \{
    2\rmi \left[-1+(1+u^2)P \right] (u^{-2}-1) \del_i
    \chi_3^{\rm CSL} \del_kP \nonumber\\
 & \hspace{1.7cm} + 2\rmi \left[-1+(1+u^{-2})P \right] (u^{2}-1) \del_i
 \chi_3^{\rm CSL} \del_kP
    \} \notag \\
    &= 2\rmi |1-u^2|^2\epsilon^{0ijk} \del_i
    \chi_3^{\rm CSL}\tr \tau_3 \del_kP \notag \\
    &= 2\rmi |1-u^2|^2\epsilon^{03jk} \del_3
   \chi_3^{\rm CSL}\tr \tau_3 \del_kP \,.
\end{align}
Inserting this expression into the second term in eq~(\ref{WZW_gauged_term}),
it becomes
\begin{gather}
    -\frac{e \mu_{\textrm{B}}}{8\pi^2}
    \epsilon^{03jk}|1-u^2|^2 \del_3
    \chi_3^{\rm CSL} A_j\del_k(\phi^{\dag}\tau_3\phi) \,.
\end{gather}
Integrating over $z$, we get
\begin{gather}
    \int_{0}^{\ell} \rmd z\, \frac{\rmi e\mu_{\textrm{B}}}{16\pi^2}\epsilon^{0ijk} A_j \tr \tau_3 (\del_i\Sigma \del_k\Sigma^{\dag} + \del_k\Sigma^{\dag}\del_i\Sigma)
    = -\frac{e \mu_{\textrm{B}}}{2\pi} \epsilon^{03jk} A_j\del_k(\phi^{\dag}\tau_3\phi) \label{WZW_second_term} \,,
\end{gather}
where we have used the integral eq.~(\ref{eq:integral_emGW_current}).
Summing up eqs.~(\ref{WZW_first_term}) and (\ref{WZW_second_term}),
the effective Lagrangian from the second term in eq.~(\ref{eq:GW}) can be  calculated as
\begin{gather}
    -\frac{e \mu_{\textrm{B}}}{2\pi}
    \epsilon^{03jk}\del_j(A_k\phi^{\dag}\tau_3\phi) \,.
\end{gather}
Finally, we arrive at the effective Lagrangian of the non-Abelian sine-Gordon soliton under the magnetic field:
\begin{align}
    \mathcal{L}_{\textrm{DW}} &\equiv \int_{0}^{\ell} \rmd z\, (\mathcal{L}_{\textrm{ChPT}} + \mathcal{L}_{\textrm{WZW}}) \notag \\
    &= - \left({\mathcal E} -\frac{e\mu_{\textrm{B}}B}{2\pi}\right) 
    + {\cal C}(\kappa)
    [(\phi^{\dag}D_{\alpha}\phi)^2
    +D_{\alpha}\phi^{\dag}D^{\alpha}\phi] \notag \\
    &+ 2\mu_{\textrm{B}}q + \frac{e \mu_{\textrm{B}}}{2\pi} \epsilon^{03jk}\del_j[A_k(1-n_3)] \,.
    \label{eq:LEET-final}
\end{align}
This is a background-gauged ${\mathbb C}P^1$ model or O(3) model with the topological terms.
Note that in the single-soliton limit ($\kappa = 1$), 
the K\"ahler class in eq.~(\ref{eq:fk}) 
reduces to ${\cal C}(\kappa=1) = 16f_\pi^2/3m_\pi^2$
recovering  our previously result \cite{Eto:2023lyo}.

\section{Domain-wall Skyrmion chain and 
domain-wall Skyrmion phase} \label{sec:DWSk-chain}
We examine Skyrmions within the domain-wall effective theory described by eq.~(\ref{eq:LEET-final}).
Initially, we neglect the gauge coupling by setting $D_{\mu} \to \partial_{\mu}$,
considering the effects of the WZW term from 
eqs.~(\ref{eq:GW}) 
and (\ref{effective_Lagrangian_GW}).
Subsequently, we incorporate the effects of the gauge coupling.
The resulting static Hamiltonian is given by
\begin{align}
    \mathcal{H}_{\textrm{DW}}
    &= 
    \frac{{\cal C}(\kappa)}{4}
    \partial_i{\bm n}\cdot \partial_i {\bm n}
    -2\mu_{\textrm{B}}q 
    -\frac{e \mu_{\textrm{B}}}{2\pi} \epsilon^{03jk}\del_j[A_k(1-n_3)]
    \label{effective_hamiltonian_without_D} \,.
\end{align}
Since the constants in eq.~(\ref{eq:LEET-final}) only give the condition of whether the domain wall appears or not, it is sufficient to consider $B>B_{\rm c}$,
and thus have been omitted in eq.~(\ref{effective_hamiltonian_without_D}). 
Then, the total energy $E_{\rm DW} = \int d^2x\, {\cal H}_{\rm DW}$ is bounded from below as
\begin{align}
    E_{\rm DW} 
    &\geq 
   2 \pi {\cal C}(\kappa) |k|
    -2\mu_{\textrm{B}}k 
    -\frac{e \mu_{\textrm{B}}}{2\pi} 
    \int d^2 x \epsilon^{03jk}\del_j[A_k(1-n_3)]
    \label{eq:BPS-bound} \,,
\end{align}
which is called the Bogomol'nyi bound,
where we have used 
\begin{eqnarray}
    \partial_i{\bm n}\cdot \partial_i {\bm n}
    = \frac{1}{2}\left(\partial_i{\bm n} \pm \epsilon_{ij}{\bm n}\times\partial_j{\bm n}\right)^2 \pm
    8 \pi q\,.
\end{eqnarray}
The inequality in eq.~(\ref{eq:BPS-bound}) 
is saturated only when the fields satisfy the (anti-)Bogomol'nyi-Prasad-Sommerfield (BPS) equation 
\cite{Polyakov:1975yp}
\begin{gather}
    \del_i\bm{n}\pm \epsilon_{ij}\bm{n}\times \del_j\bm{n}=0 \,,
\end{gather}
where the upper (lower) sign corresponds to the (anti-)BPS 
equation.
It is interesting to observe that 
the second term in eq.~(\ref{effective_hamiltonian_without_D})
splits energies between BPS lumps $k>0$ and 
anti-BPS lumps $k<0$~\footnote{
This situation is analogous to magnetic Skyrmions 
in chiral magnets (see, e.g. ref.~\cite{Ross:2020hsw}), 
in which case the so-called 
Dzyaloshinskii–Moriya interaction plays 
such a role.
}. 
The BPS solutions to this equation
characterized by the winding number $k$ ($>0$)
is given by \cite{Polyakov:1975yp}
\begin{eqnarray}
        n_3 = \frac{1-|f|^2}{1 + |f|^2}, \quad
        f = \frac{b_{k-1}w^{k-1}+\cdots+b_0}{w^k + a_{k-1}w^{k-1}+\cdots+a_0}\,,
        \label{sol_n}
\end{eqnarray}
where $w \equiv x+\rmi y$, and the set of complex parameters $\{a_A,b_A\}$ ($A=0,1,\cdots, k-1$)
are the moduli parameters.

We examine the gauge coupling between $\bm{n}$ and $A_\alpha$. With the electromagnetic gauge symmetry $U(1)_{\rm EM}$ generated by $\tau_3$, the transformation of $n_1 + \rmi n_2$ is given by 
\begin{equation}
n_1 + \rmi n_2 \to \rme^{-\rmi\lambda}(n_1+\rmi n_2),
\end{equation}
while $n_3$ remains neutral. The covariant derivative is
\begin{equation}
D_\alpha (n_1+\rmi n_2) = (\partial_\alpha - \rmi e A_\alpha)(n_1+\rmi n_2).
\end{equation}
Let $C$ be a closed curve where $n_3 = 0$ and $D$ its interior. Due to the spontaneous breaking of $U(1)_{\rm EM}$ symmetry around $|n_1+\rmi n_2|=1$, curve $C$ functions as a superconducting loop, carrying a persistent current. Expressing $n_1+\rmi n_2 = \rme^{i\psi}$ on $C$, the gauge field configuration along $C$ is determined by the gradient energy minimization, leading to $|D_\alpha (n_1+\rmi n_2)|^2 = 0$ and $\partial_\alpha \psi = e A_\alpha$. Consequently, the flux and area quantization on $D$ becomes
\begin{equation}
B S_D = \int_D \rmd^2x\, B = \oint_C \rmd x^i A_i = \frac{1}{e} \oint_C \rmd x^i \partial_i \psi = \frac{2\pi k}{e},
\label{quantize}
\end{equation}
where $k$ is the lump number on $D$ and $S_D$ is the area of $D$.

For a single lump with $k=1$, represented by \(f = \frac{b_0}{w}\), the size and phase moduli are \(|b_0|\) and \(\text{arg } b_0\), respectively. The relationship for \(n_3\) is 
\begin{equation}
n_3 = \frac{|w|^2-|b_0|^2}{|w|^2+|b_0|^2},
\end{equation}
and the region size $D$ defined by \(n_3=0\) is \(|w|=|b_0|\). The flux quantization requires the size modulus to be 
\begin{equation}
    k=1: \quad |b_0|=\sqrt{\frac{2}{eB}} .\label{eq:flux-quant-k=1}
\end{equation}
For axially symmetric \(k\)-lumps, with \(f = \frac{b_0}{w^k}\), \(n_3\) is 
\begin{equation}
n_3 = \frac{|w|^{2k}-|b_0|^2}{|w|^{2k}+|b_0|^2},
\end{equation}
and the flux quantization requires the size modulus to be 
\begin{equation}
 |b_0|=\left(\frac{2k}{eB}\right)^{k/2} .    
\end{equation}
%
\begin{figure}[tp]
  \begin{center}
   \includegraphics[width=15.0cm]{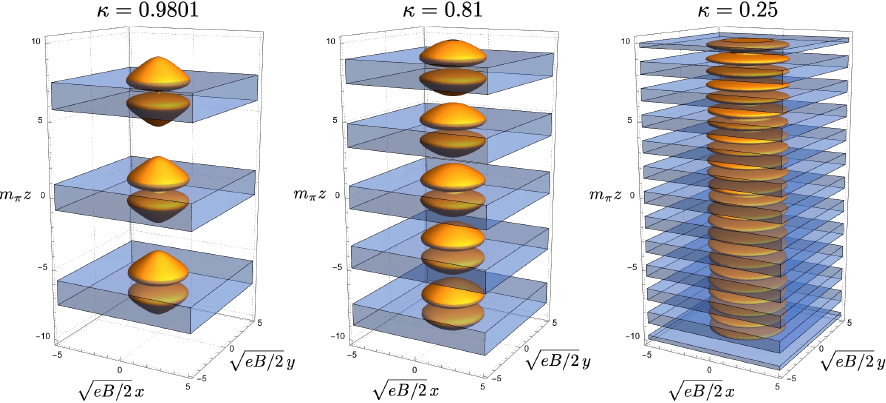}
  \end{center}
  \caption{
   The minimal $(k=1)$ Skyrmion chain in the CSL with $f(w) = \sqrt {2/eB}/w$. 
   The $z$ coordinate is rescaled by $m_{\pi}$ as $m_{\pi} z$, 
   and the $x$ and $y$ coordinates are rescaled by the lump size 
   $\sqrt {2/eB}$ as  $\sqrt{2/eB} (x,y)$.
   The isosurface of baryon number density ${\cal B} = 1/(10\pi^2)$ (orange), and the sine-Gordon soliton $\pi/2 < \theta < 3\pi/2$ (blue). 
   From the left to right panels, the elliptic modulus $\kappa$ decreases ($B \mu_{\rm B}$ increases)  
   and the periodicity of the CSL decreases.
   The size modulus is fixed by the quantization condition, 
   the physical width of the right  configuration is smaller than that of the left configuration 
   since smaller $\kappa$ corresponds to larger $B$.
   In the  rescaled coordinates, the shape changes from round (left) to crushed (right), 
   and they look like periodic macarons (left) to a pancake tower (right).
}
  \label{fig:dwsk_chain}
\end{figure}

Since Skyrmions sit in the same positions 
$(x^1,x^2)$ at each soliton, the total configuration in 3+1 dimensions is 
domain-wall Skyrmion chains.
In fig.~\ref{fig:dwsk_chain}, we plot our solutions of domain-wall Skyrmion chains 
for various elliptic modulus $\kappa$.
 The orange regions denote the isosurface of the baryon number density ${\cal B} = 1/(10\pi^2)$, and 
 the blue regions denote the soliton $\pi/2 < \theta < 3\pi/2$. 
One can confirm that one lump on one soliton 
is composed of two Skyrmions as can be expected from 
eq.~(\ref{eq:pi2-pi3}).
From the left to right panels, the elliptic modulus $\kappa$ decreases 
corresponding to the situation that 
$B \mu_{\rm B}$ increases, 
   and the periodicity of the CSL decreases.
   From left to right, these look like
from periodic macarons to
a pancake tower.
We note that a similar domain-wall skyrmion chain 
has been studied in chiral magnets in $2+1$ simensions \cite{Amari:2023bmx}.

Now let us discuss a constraint from the second term of WZW term 
following ref.~\cite{Eto:2023lyo}.
The integration of the last term in eq.~(\ref{eq:BPS-bound}) can be rewritten as 
\begin{eqnarray}
 - \int d^2 x \frac{e \mu_{\textrm{B}}}{2\pi} \epsilon^{03jk}\del_j[A_k(1-n_3)]
 = \frac{
    e \mu_{\textrm{B}}B}{4\pi}\oint \rmd S_i\,
    x^i (n_3-1) 
    =  e\mu_{\textrm{B}}B |b_{k-1}|^2\
    \,,    
\end{eqnarray}
where 
we have used the explicit solution in eq.~(\ref{sol_n}) 
in the last expression. 
We thus reach 
the energy of domain-wall Skyrmions, given by
\begin{gather}
    E_{\rm DWSk} = 
    2 \pi {\cal C}(\kappa) |k|
    -2\mu_{\textrm{B}}k
    + e\mu_{\textrm{B}}B |b_{k-1}|^2\,.
    \label{E_3} 
\end{gather}
For a single lump ($k=1$), 
we have $E_{\rm DWSk} = 2 \pi {\cal C}(\kappa)$ with the cancellation 
between the second and third terms due to
the flux quantization 
in eq.~(\ref{eq:flux-quant-k=1})
which is always positive. 
For higher winding $k\geq 2$, 
we have a further constraint 
\begin{equation}
  b_{k-1} = 0,    
\end{equation}
to minimize the domain-wall Skyrmion energy in eq.~(\ref{E_3}), which 
can become  negative for sufficiently large $\mu_{\rm B}$ from its second term.

Finally, let us discuss the DWSk phase in which 
Skyrmions are created spontaneously.
From the above consideration, 
the phase boundary between 
the CSL and DWSk phases is determined to be
\begin{eqnarray}
    \mu_{\rm c}  = \pi {\cal C}(\kappa) 
    = \frac{16 \pi f_{\pi}^2}{3m_{\pi}} 
    \frac{(2-\kappa^2) E(\kappa) 
    - 2 (1-\kappa^2) K(\kappa)}{\kappa^3}.
    \label{eq:phase-boundary}
\end{eqnarray}
When $\mu_{\rm B} \geq \mu_{\rm c}$, 
the lumps have negative energy 
and are spontaneously created, implying 
the DWSk phase. 
Note that in the single-soliton limit 
$\kappa =1$, $\mu_{\rm c}$ in eq.~(\ref{eq:phase-boundary}) 
reduces to a constant 
 $\frac{16 \pi f_{\pi}^2}{3m_{\pi}}$ 
 (thus a vertical line)
 reproducing the previous result in 
 eq.~(\ref{eq:negative}) \cite{Eto:2023lyo}. 
Since the elliptic modulus $\kappa$ is determined from 
$B$ and $\mu_{\rm B}$ in general, 
eq.~(\ref{eq:phase-boundary})
gives a nontrivial curve 
 beyond the one-soliton approximation, 
represented by the red curve in 
 fig.~\ref{fig:phase_diagram}. 
 This is our main result.
It is interesting to observe 
that $\mu_{\rm c}$ can be interpreted as 
the effective nucleon mass in this medium 
(inside solitons with 
the chemical potential $\mu_{\rm B}$
and magnetic field $B$), 
which is $\frac{16\pi f_{\pi}^2}{3m_{\pi}}
    \sim 1.03 \;\; {\rm GeV}$ 
    at the tricritical point 
    (the white dot in fig.~\ref{fig:phase_diagram}), 
    and it becomes lighter as the magnetic field is stronger.

Let us compare this boundary with 
the instability curve of the CSL configuration
via the charged pion condensation in eq.~(\ref{eq:instability})
by Brauner and Yamamoto 
\cite{Brauner:2016pko}.
The full expression for the instability curve is determined by eliminating the elliptic modulus $\kappa$ from the following two equations  
\cite{Brauner:2016pko} 
\begin{eqnarray}
    && B_{\rm CPC} = \frac{m_{\pi}^2}{\kappa^2}
    \sqrt{1-\kappa^2+\kappa^4} \,, 
    \nonumber \\
    && \frac{E(\kappa)}{\kappa} = \frac{e\mu_{\mathrm{B}}
    B_{\rm CPC}}{16\pi m_{\pi}f_{\pi}^2} ,\, 
    \label{eq:instability-full}
\end{eqnarray}
and is denoted by the green dotted curve in fig.~\ref{fig:phase_diagram}. 
One can observe that this curve is entirely above the phase boundary between 
 the CSL and DWSk phases 
in eq.~(\ref{eq:phase-boundary}), denoted by the red curve in fig.~\ref{fig:phase_diagram}. 
The CSL configuration remains 
 locally stable (metastable)  
 in the region between the red and green dotted curves.

Here, let us investigate the large $B$ behaviours of these two curves. 
To this end, we expand the equations around $\kappa=0$. 
Expanding eqs.~(\ref{eq:phase-boundary})
and (\ref{eq:minimization_condition}) 
 around $\kappa=0$,
we obtain  
\begin{eqnarray}
  &&  \mu_{\rm B} \sim \frac{\pi f_{\pi}^2 \kappa}{2 m_{\pi}},\\
  &&  \kappa \sim \frac{8\pi^2 m_{\pi} f_{\pi}}{\mu_{\rm B} B_{\rm c}},
\end{eqnarray}
respectively. 
Eliminating $\kappa$ from these two equations,  
we obtain 
\begin{eqnarray}
    B_{\rm c} \sim \frac{4 \pi^3 f_{\pi}^4}{\mu_{\rm B}^2} .\label{eq:asympt}
\end{eqnarray}
On the other hand, 
expanding the instability curve of 
eq.~(\ref{eq:instability-full}) in the same way, 
we find that it asymptotically behaves as in eq.~(\ref{eq:instability})
\begin{eqnarray}
    B_{\rm CPC} 
    \sim \frac{16 \pi^4 f_{\pi}^4}{\mu_{\rm B}^2} 
    = 4\pi B_{\rm c}.
\end{eqnarray}
Clearly, this is above the phase boundary 
between the CSL and DWSk phases in eq.~(\ref{eq:asympt}).

\section{Summary and discussion}
\label{sec:summary}

In this paper, in the phase diagram of QCD with finite baryon density and magnetic field, we have determined  the phase boundary between 
the CSL and DWSk phases beyond the single-soliton approximation 
at the leading order ${\cal O}(p^2)$ of ChPT. 
The key point to go beyond 
the single-soliton approximation 
is considering domain-wall Skyrmion chains  
in multiple soliton configurations.
We have constructed the low-energy effective theory 
of one period of the CSL 
by the moduli approximation. 
We have obtained 
in eq.~(\ref{eq:LEET-final}) 
the background-gauged ${\mathbb C}P^1$ model or O(3) model with  topological terms originated from 
the WZW term, topological lump charge 
 in 2+1 dimensional worldvolume 
and the topological term for the soliton number.
A single topological lump in 
 2+1 dimensional worldvolume theory 
 is a superconducting ring. 
 Due to the flux quantization condition 
in eq.~(\ref{quantize}), the size modulus is fixed.
 We have determined 
 the phase boundary between 
 the CSL and DWSk phases in eq.~(\ref{eq:phase-boundary}) denoted by
 the red curve in fig.~\ref{fig:phase_diagram}
 from the negative energy condition of the lumps.
 We have found that a large region in the CSL phase is occupied 
 by the DWSk phase, and that 
  the CSL configuration is metastable in the region between 
 the red curve and green dotted curve given by eq.~(\ref{eq:instability-full}), 
 beyond which the CSL is unstable.
 The blue, red and green dotted curves meet at the tricritical point in eq.~(\ref{eq:TCP}).

We have worked out at the leading order ${\cal O}(p^2)$ of the ChPT for which we have not needed higher derivative terms such as the Skyrme term. At this order, the magnetic field is a background field.
At the next leading order 
${\cal O}(p^4)$, one needs higher derivative terms as well as the kinetic term of the electromagnetic gauge field. 
  The stability beyond the leading order remains a future problem.

The phase transition between 
 the CSL and DWSk phases in eq.~(\ref{eq:phase-boundary}) denoted by
 the red curve in fig.~\ref{fig:phase_diagram} 
 would be the so-called second order of nucleation type in the classification by de Gennes \cite{deGennes1975}
 In such a case, the configuration of one side of the boundary often remains metastable on the other side, which is in fact our case. Similarly, the phase boundary between the QCD vacuum and CSL denoted by the blue curve in fig.~\ref{fig:phase_diagram} was recently shown to be of the second order 
 \cite{Brauner:2023ort}, 
 and it should be of the nucleation type. 
 Quantum nucleation, explored in the case of the transition from the vacuum to CSL  \cite{Eto:2022lhu,Higaki:2022gnw}, 
 should be applied to the transition from the CSL to DWSk phase.
 Investigating this transition is one of important future directions.

In this paper, we have obtained the periodic structure of Skyrmions:  
a domain-wall Skyrmion chain. 
Another approach is to construct 
an effective theory of a soliton lattice.
The effective theory of each soliton is a  ${\mathbb C}P^1$ model or O(3) model. 
As was studied for a non-Abelian vortex lattice in ref.~\cite{Kobayashi:2013axa}, we can construct a lattice effective theory as follows.
A neighboring pair of solitons interacts 
as $H_{\rm int} = - J \sum_{\left<i,i+1\right>}  {\bf n}_i {\bf n}_{i+1}$ 
with the ${\mathbb C}P^1$ moduli 
${\bf n}_i$ of the $i$-th soliton.
In our case, the lattice behaves as a ferromagnet with $J >0$, 
and thus the moduli tend to be aligned. 
When one constructs a lump on a soliton, 
the system prefers to place the same lumps 
on its neighboring solitons. 
 Then, that theory admits an array of 
 lumps along the lattice direction, which is nothing but our Skyrmion chain.
One can also take a continuum limit (large $B \mu_{\rm B}$) 
resulting in a 3+1 dimensional anisotropic ${\mathbb C}P^1$ model , in which we need a careful treatment for the terms from the WZW term. 
Then, the continuum theory  should admit 
a lump string along the $z$-direction, 
which should have negative energy.

Let us discuss a possible 
relation between 
our configuration of the domain-wall Skyrmion chain and an Abrikosov vortex lattice in the charged pion condensation
proposed in Ref.~\cite{Evans:2022hwr}. 
It was shown 
in Refs.~\cite{Nitta:2015mxa,Nitta:2022ahj}
that 
when (ungauged) Skyrmions are periodically arranged with a twisted boundary condition, they reduce to 
global vortices in the small periodicity limit.\footnote{
This relatoon is analogous to those of 
periodic Yang-Mills instantons (Calorons) 
with a twisted boundary condition that reduce to 
monopoles in the same limit 
\cite{Lee:1998vu,Lee:1998bb,Kraan:1998pm,Kraan:1998sn}, 
and  
periodic ${\mathbb C}P^{N-1}$ lumps with a twisted boundary condition that reduce to 
${\mathbb C}P^{N-1}$ domain walls in the same limit \cite{Eto:2004rz,Eto:2006mz,Eto:2006pg}. 
Such relations are known as a T-duality in string theory context.
Moreover, these three relations are 
actually related to each other as follows.
SU($N$) Yang-Mills intantons become SU($N$) 
Skyrmions \cite{Eto:2005cc} 
inside a non-Abelian domain wall \cite{Shifman:2003uh,Eto:2008dm} 
whose worldvolume effective theory is 
the Skyrme model \cite{Eto:2005cc}, 
and they also become ${\mathbb C}P^{N-1}$ 
lumps \cite{Eto:2004rz,Hanany:2004ea}
inside a non-Abelian vortex 
whose worldsheet effective theory is 
the ${\mathbb C}P^{N-1}$ model \cite{Hanany:2003hp,Auzzi:2003fs,Eto:2005yh}.
}
See the rightmost panel of Fig.~\ref{fig:dwsk_chain}. 
In our case, these vortices should carry baryon numbers 
\cite{Gudnason:2014hsa,Gudnason:2016yix,Nitta:2015tua}. 
This  may offer 
a possible crossover between 
our configuration of the Skyrmion chain and an Abrikosov vortex lattice  \cite{Evans:2022hwr}.
However, there is a significant difference.
If we turn on a dynamical electromagnetic 
gauge field
at the next leading order 
${\cal O}(p^4)$, they would reduce to 
superconducting strings 
since charged pions are condensed in the vortex cores. 
Thus, our Skyrmion chains at the next leading order 
${\cal O}(p^4)$ are superconducting strings 
in the short period limit
(that is the continuum limit 
of the Heisenberg spin chain at large $B \mu_{\rm B}$ as mentioned above).

Before concluding this paper, we wish to comment on a domain-wall Skyrmion phase analogous to the one found in QCD matter under rapid rotation~\cite{Eto:2023tuu}. Recent years have seen a surge in interest regarding rotating QCD matter~\cite{Chen:2015hfc,Ebihara:2016fwa,Jiang:2016wvv,Chernodub:2016kxh,Chernodub:2017ref,Liu:2017zhl,Zhang:2018ome,Wang:2018zrn,Chen:2019tcp,
Chernodub:2020qah,Chernodub:2022veq,
Chen:2021aiq,Huang:2017pqe,Nishimura:2020odq,Eto:2021gyy}, primarily due to the observation of an exceptionally large vorticity of the order of $10^{22}/{\rm s}$ in quark-gluon plasmas produced in non-central heavy-ion collision experiments at the Relativistic Heavy Ion Collider (RHIC)~\cite{STAR:2017ckg,STAR:2018gyt}. 
In ChPT, the anomalous term for the $\eta'$ meson was derived in~\cite{Huang:2017pqe,Nishimura:2020odq} by matching it with the chiral vortical effect (CVE)~\cite{Vilenkin:1979ui,Vilenkin:1980zv,Son:2009tf,Landsteiner:2011cp,Landsteiner:2012kd,Landsteiner:2016led} in the context of mesons. Analogous to the effect of a magnetic field, this term suggests a CSL composed of the $\eta'$ meson during rapid rotation~\cite{Huang:2017pqe,Nishimura:2020odq,Chen:2021aiq}. For two flavor scenarios, the phenomenon manifests as an $\eta$-CSL made up of the $\eta$ meson. In a significant parameter region, a single $\eta$-soliton energetically decays into a couple of non-Abelian solitons, leading to neutral pion condensation in its vicinity. 
A lone non-Abelian soliton breaks the vector symmetry ${\rm SU}(2)_{\rm V}$ down to its U$(1)$ subgroup. This results in NG modes described by ${\rm SU}(2)_{\rm V}/{\rm U}(1) \simeq {\mathbb C}P^1 \simeq S^2$, which localize near the soliton as discussed in~\cite{Eto:2021gyy}. Therefore, mirroring the $\pi_0$ soliton in a magnetic setting, each non-Abelian soliton carries ${\mathbb C}P^1$ moduli and is termed a non-Abelian sine-Gordon soliton~\cite{Nitta:2014rxa,Eto:2015uqa,Nitta:2015mma,Nitta:2015mxa,Nitta:2022ahj}. Relying on the single-soliton approximation, we posited the DWSk phase for rapid rotations in~\cite{Eto:2023tuu}. Consequently, our present study on a domain-wall Skyrmion chain based on multiple solitons can be extended to rotational scenarios.

\begin{acknowledgments}
This work is supported in part by 
 JSPS KAKENHI [Grants No. JP19K03839 (ME) 
No. JP21H01084 (KN), 
 and No. JP22H01221 (ME and MN)] and the WPI program ``Sustainability with Knotted Chiral Meta Matter (SKCM$^2$)'' at Hiroshima University (KN and MN).
\end{acknowledgments}

\bibliographystyle{jhep}
\bibliography{reference.bib}


\end{document}